\documentclass[12pt]{article}
\usepackage{mathrsfs}
\usepackage{amsfonts}
\usepackage{palatino}
\usepackage{textcomp}
\usepackage{color}
\usepackage{amsmath}
\usepackage{graphicx}
\usepackage{amssymb}
\usepackage{authblk}
\usepackage{float}

\renewcommand{\theequation}{\arabic{section}.\arabic{equation}}

\begin{document}

\title{\bf Bethe Ansatz for XXX chain\\
 with negative spin}
\date{}

\author[1,2]{\large \bf Kun Hao\footnote{haoke72@163.com}}
\author[3,4]{\large \bf Dmitri Kharzeev}
\author[2]{\large \bf Vladimir Korepin}

\affil[1]{\normalsize Institute of Modern Physics, Northwest University, Xi'an 710069, China}
\affil[2]{\normalsize C.N. Yang Institute for Theoretical Physics, Stony Brook University, NY 11794, USA}

\affil[3]{\normalsize Department of Physics and Astronomy, Stony Brook University, NY 11794, USA}

\affil[4]{\normalsize Physics Department and RIKEN-BNL Research Center,

Brookhaven National Laboratory, Upton, NY 11973, USA}


\maketitle
\vspace{-0.5cm}

XXX spin chain with spin $s=-1$ appears as an effective theory of  Quantum Chromodynamics.
It is equivalent to lattice nonlinear Schroediger's equation: interacting chain of  harmonic oscillators [bosonic].
In thermodynamic limit each energy level is a scattering state of several elementary excitations [lipatons].
Lipaton is a fermion: it can be represented  as a topological  excitation [soliton] of original [bosonic] degrees of freedom, described by  the group $Z_2$  .
We also provide the CFT description (including local quenches) and Yang-Yang thermodynamics of the model.

\section{Introduction}
\setcounter{equation}{0}
In a ground-breaking paper \cite{Lipatov94}, Lev Lipatov discovered that high-energy Quantum Chromodynamics (QCD) can be described by the XXX spin chain with spin $s=0$.
He also found corresponding eigenstates describing the high energy behavior of the scattering amplitudes. This behavior is consistent with the high energy, small Bjorken $x$ behavior of the structure functions measured in deep-inelastic scattering (DIS).

At high energy, the scattering amplitudes in QCD are described by the exchange of gluons dressed by virtual gluon loops -- so-called reggeized gluons. In the limit of large number of colors $N_c$ (with fixed $g^2 N_c$, where $g$ is the QCD coupling), the corresponding Feynman diagrams have the topology of the cylinder. Therefore the Hamiltonian describing the interactions of reggeized gluons reduces to the sum of terms describing the pairwise near-neighbor interactions, similar to the Hamiltonian of a spin chain.

 L. D. Faddeev and G. P. Korchemsky in   \cite{Faddeev-Korchemsky1995} mapped the spin $0$ model to spin $-1$.  They also diagonalized the  XXX Hamiltonian by algebraic Bethe Ansatz \cite{ABA,Korepin93}.

Recently, it has been argued that the structure functions measured in DIS can be interpreted in terms of entanglement entropy between the spatial region probed by the virtual photon and the rest of the nucleon \cite{Kharzeev:2017qzs}. To develop this description further and to describe the real-time evolution of the final state in DIS, one needs to identify the physical excitations of the effective high energy QCD Hamiltonian. This can be conveniently done in the thermodynamic limit at zero temperature, when both the number of reggeized gluons $L$ and the number $N$ go to infinity, with a fixed ratio $N/L$. This motivates our study of thermodynamics (local quenches and modeling the deep-inelastic collisions) in the XXX spin chain with negative spin.

The XXX chain with negative spin can be considered as discretization of nonlinear Schroedinger equation (NLS) \cite{Izergin-Korepin1982,Korepin-Izergin1982,Korepin93}.
The thermodynamics Bethe Ansatz method is based on C. N. Yang and C. P. Yang's work \cite{Yang1968}, in which they considered thermodynamics of continuous NLS.

The algebraic Bethe Ansatz (quantum inverse scattering method) \cite{ABA,Korepin93} can be applied to the solution
of equations for wave functions of compound states
which describe $L$ reggeized
gluons in the multicolor QCD in a generalized leading logarithmic
approximation (LLA).
In the following, we will first briefly introduce the connection of QCD in the LLA limit to the XXX spin chain for spin $s=-1$, and the relation between XXX spin chain with negative spin and quantum nonlinear Schroedinger model.
These relations enable us to further investigate the thermodynamics of the models.

The holomorphic multicolor QCD Hamiltonian \cite{Lipatov94,Faddeev-Korchemsky1995} describes the nearest neighbor interactions of $L$ particles (reggeized gluons):
\begin{eqnarray}
{\cal H} = \sum_{k=1}^L H_{k,k+1},
\label{H-gen}
\end{eqnarray}
with periodic boundary conditions $H_{L,L+1}=H_{L,1}$. The local Hamiltonians are given by the equivalent representations\footnote{Note that the difference between this Hamiltonian and the one proposed by Lipatov in \cite{Lipatov94} is a minus sign.}
\begin{eqnarray}
H_{j,k}
&=&-P_j^{-1}\ln(z_{jk})P_j
           -P_k^{-1}\ln(z_{jk})P_k
           -\ln(P_jP_k)-2\gamma_E
\nonumber
\\
&=&
-2\ln(z_{jk})-(z_{jk})\ln(P_jP_k)(z_{jk})^{-1}-2\gamma_E,
\label{H2}
\end{eqnarray}
where $P_j=i{\partial/\partial z_j}=i\partial_j$, $z_{jk}=z_j-z_k$, and $\gamma_E$ is the Euler constant. We have to put $j=k+1$ and substitute into (\ref{H-gen}).

Lipatov used holomorphic representation of $su(2)$
\begin{eqnarray}
S_k^+ = z_k^2\partial_k-2s z_k ,
\qquad
S_k^- = -\partial_k ,
\qquad
S_k^z = z_k \partial_k - s,
\qquad k=1,\cdots,L.
\label{S}
\end{eqnarray}
He mapped DIS to a spin chain.
The definition of the chain is based on the existence of a fundamental matrix $R_{jk}(\lambda)$
which obeys the Yang-Baxter equation
\begin{eqnarray}
R_{jk}(\lambda)=f(s,\lambda)
\frac{\Gamma(i\lambda-2s)\Gamma(i\lambda+2s+1)}
                         {\Gamma(i\lambda-J_{jk})\Gamma(i\lambda+J_{jk}+1)}.
\end{eqnarray}
Here $f(s,\lambda)$ is a complex valued function (it normalizes the $R$ matrix). The $\lambda$ is called spectral parameter; we shall express everything as a function of $\lambda$ (including the energy and momentum of lipaton). The operator $J_{jk}$ is defined in the space $V\otimes V$ as a solution of the operator equation,
\begin{eqnarray}
J_{jk}(J_{jk}+1) = 2\vec S_j\otimes\vec S_k.
\label{J-operator}\end{eqnarray}
Then the Hamiltonian of the XXX model with spin $s=0$ describes interaction of nearest neighbors, see (\ref{H-gen})
\begin{eqnarray}
H_{jk}=\left.\frac1{i}\frac{d}{d\lambda}
    \ln R^{(s=0)}_{jk}(\lambda)\right\vert_{{}_{\lambda=0}},\quad
    H_{jk}=-\psi(-J_{jk})-\psi(J_{jk}+1)+2\psi(1).
\label{H2xxx}
\end{eqnarray}
Here $\psi(x)=d \ln\Gamma(x)/dx$, and $\psi(1)=-\gamma_E$ ($\gamma_E$ is the Euler constant). The operator $J_{jk}$ is a solution of (\ref{J-operator}) when $s=0$,
\begin{eqnarray}
J_{jk}(J_{jk}+1)=-(z_j-z_k)^2\partial_j\partial_k.
\label{spin-J}\end{eqnarray}
Put $j=k+1$ in (\ref{H2xxx}).
This is a description of DIS in QCD by $s=0$ spin chain.
L. D. Faddeev and G. P. Korchemsky mapped the spin $0$ chain to spin $-1$.
The similarity transformation was found in \cite{Faddeev-Korchemsky1995}.
\begin{equation}
{\cal H}^{(s=-1)}=(z_{12}z_{23}\cdots{z}_{L1})^{-1}{\cal H}^{(s=0)}z_{12}z_{23}\cdots{z}_{L1}.
\end{equation}
The $s=-1$ Hamiltonian has $4$-nearest neighbor interactions.
The XXX model with spin $s=-1$ was solved by the algebraic Bethe Ansatz in \cite{Faddeev1980,Tarasov-Takhtajan-Faddeev1983}.

The corresponding Bethe equations for $s=-1$ determine the parameters $(\lambda_1,\ldots,\lambda_N)$:
\begin{eqnarray}
\left({\frac{\lambda_k+is}{\lambda_k-is}}\right)^L
=\prod_{j=1\atop j\neq k}^N \frac{\lambda_k-\lambda_j+i}{\lambda_k-\lambda_j-i}\xrightarrow{s=-1}\left({\frac{\lambda_k-i}{\lambda_k+i}}\right)^L
=\prod_{j=1\atop j\neq k}^N \frac{\lambda_k-\lambda_j+i}{\lambda_k-\lambda_j-i},\;\,
k=1,\cdots,N.
\label{Beq}
\end{eqnarray}
These are periodic boundary conditions. Later we argue that in order to construct elementary excitation, we have to change to anti-periodic boundary conditions.
We show in Appendix A that all the solutions $\lambda_k$ of the above Bethe equations are real numbers. This means that there is \textbf{no bound state in this system}. The following thermodynamics analysis is based on this.

The explicit expressions for the eigenvalues of integrals of motions
for arbitrary spin $s$ have been found in algebraic Bethe Ansatz \cite{Tarasov-Takhtajan-Faddeev1983}
and we use these expressions for $s=-1$ to get the eigenvalues of the Hamiltonian
\begin{equation}
E\equiv\sum_{j=1}^N\frac1{i}\frac{d}{d\lambda_j}
\ln\frac{\lambda_j+i}{\lambda_j-i}=\sum_{j=1}^N \frac{-2}{\lambda^2_j+1},
\label{energy}\end{equation}
where $\{\lambda_j\}$ obey the Bethe equations (\ref{Beq})
for a fixed number of reggeized gluons $L$. Thus, this relation yields the spectrum of  original holomorphic QCD model with Hamiltonian $\cal H$.

The quantum lattice nonlinear Schroedinger's equation was introduced in \cite{Izergin-Korepin1982,Korepin-Izergin1982,Korepin93}. We call it NLS. It is equivalent to XXX spin chain with negative spin.
Quantum lattice NLS is a chain of interacting harmonic oscillators: Let $\Psi_j^*$ and $\Psi_k$ be the canonical creation and annihilation operators of the harmonic oscillator:
\begin{equation}
[\Psi_j,\Psi_k^*]=\delta_{jk},
\end{equation}
and
\begin{equation}
\varrho_j=(1+{\kappa\Delta\over4}\Psi_j^*\Psi_j)^{1\over2},
\end{equation}
here $\kappa>0$ is coupling constant\footnote{$\kappa$ is also a function of spin $s$: $\kappa=\big|{2/{s\Delta}}\big|$. We will analyze the strong coupling case $\kappa\rightarrow\infty$ later.} for NLS and $\Delta>0$ is a step of the lattice.
The operators
\begin{equation}
{S_j}^x={i\over\sqrt{\kappa\Delta}}(\Psi_j^*\varrho_j+\varrho_j\Psi_j),\quad
{S_j}^y={1\over\sqrt{\kappa\Delta}}(\varrho_j\Psi_j-\Psi_j^*\varrho_j),\quad
{S_j}^z={-2\over\kappa\Delta}(1+{\kappa\Delta\over2}\Psi_j^*\Psi_j),
\end{equation}
are the generators of an irreducible representation of $SU(2)$ algebra with arbitrary negative spin
\begin{equation}
 s=-\frac{2}{\kappa\Delta}.
 \label{spinValue}\end{equation}
In general, this $SU(2)$ representation \cite{Izergin-Korepin1982,Tarasov-Takhtajan-Faddeev1983} is infinite-dimensional, but for special (negative) values of $\Delta$ it is finite-dimensional.

Let us focus on the correspondence between Bethe equations of the two models.
The Bethe roots $\lambda_{k}$ of quantum lattice NLS model satisfy the following Bethe equations,
\begin{equation}
\left(\frac{1+i\lambda_{k}\Delta /2}{1-i\lambda_{k}\Delta /2} \right)^{L}=\prod^N_{j\neq k} \frac{\lambda_{k}-\lambda_{j}+i\kappa}{\lambda_{k}-\lambda_{j}-i\kappa}\;
\Longleftrightarrow\left(  \frac{i({-2\over{\kappa}\Delta})\kappa+\lambda_{k}}{i({-2\over{\kappa}\Delta})\kappa-\lambda_{k}} \right)^{L}=\prod^N_{j\neq k} \frac{\lambda_{k}-\lambda_{j}+i\kappa}{\lambda_{k}-\lambda_{j}-i\kappa}.
\label{Beq-NLS}
\end{equation}
Comparison of the above modified Bethe equations and the Bethe equations (\ref{Beq}) shows the connections between the two models.
When we take coupling constant $\kappa=1$, and $\Delta=2$, the Bethe equations become
\begin{equation}
\left(  \frac{i({-2\over{\kappa}\Delta})\kappa+\lambda_{k}}{i({-2\over{\kappa}\Delta})\kappa-\lambda_{k}} \right)^{L}=\prod^N_{j\neq k} \frac{\lambda_{k}-\lambda_{j}+i\kappa}{\lambda_{k}-\lambda_{j}-i\kappa}
\,\xrightarrow{\kappa=1,\Delta=2}\,
(-1)^{L}\left(\frac{\lambda_{k}-i}{\lambda_{k}+i}\right)^{L}=\prod^{N}_{j\neq k}\frac{\lambda_{k}-\lambda_{j}+i}{\lambda_{k}-\lambda_{j}-i}.
\label{Bethe-eq-NLS}
\end{equation}

This means that quantum lattice NLS model describes a more general XXX spin chain model with negative spin $s=-2/\kappa\Delta$, and holomorphic QCD is its special case with spin $s=-1$, $\Delta=2$ and coupling constant $\kappa=1$.

In order to analyze the Bethe roots distribution for negative spin $s$, first we take the logarithm of Bethe equations (\ref{Beq-NLS}) for quantum lattice NLS,
\begin{equation}
2\pi n_k=\sum^N_{j=1}\theta(\lambda_k-\lambda_j)+L\;\theta\left({\lambda_k/|s|}\right),
\label{Log-Bethe-eq-NLS}\end{equation}
where
\begin{eqnarray}
&&\theta(\lambda)=-\theta(-\lambda)=i\ln\left(\frac{i\kappa+\lambda}{i\kappa-\lambda}\right);\quad
-\pi<\theta(\lambda)<\pi,\quad \text{Im\,}\lambda=0,\quad\kappa>0.
\end{eqnarray}
This function is monotonically increasing,
\begin{equation}
\theta'(\lambda-\mu)=K(\lambda,\mu)=\frac{2\kappa}{\kappa^2+(\lambda-\mu)^2},\qquad
K(\lambda)=K(\lambda,0).
\end{equation}
$\kappa=-2/s\Delta=\big|2/s\Delta\big|$ is coupling constant. Spin $s=-2/\kappa\Delta$ is negative (with $\kappa$ and $\Delta$ being arbitrary positive numbers). In this paper,
we mainly consider $s=-1$ (corresponding to $\Delta=2$ and $\kappa=1$), unless specified otherwise.

The logarithmic form of Bethe equations (\ref{Beq}) for holomorphic QCD model (XXX with spin $s=-1$) is:
\begin{equation}
2\pi n_k=\sum^N_{j=1}\theta(\lambda_k-\lambda_j)+L\;\theta(\lambda_k).
\label{Log-Bethe-eq}\end{equation}
Corresponding to the solutions of the Bethe equations (\ref{Log-Bethe-eq}), the energy function (\ref{energy}) $-\sum_{j=1}^N{2/(\lambda^2_j+1)}$ with fixed number $N$ is minimized by the sequential set of integer ($N$ odd) or half-integer ($N$ even) numbers $n_j$:
\begin{equation}
n_j=-\left(\frac{N-1}{2}\right)+j-1,\quad j=1,\cdots,N.
\label{nj}\end{equation}
So the set of integers for the ground state in sequential $(n_{j+1}-n_j=1)$. It fills the interval from $-(N-1)/2$ to $(N-1)/2$.

The above logarithmic equations (\ref{Log-Bethe-eq-NLS}) and (\ref{Log-Bethe-eq})
show the Bethe roots distributions.
The results further display that quantum lattice NLS model is a generalization of XXX model with negative spin. Holomorphic QCD model (XXX spin chain with spin $s=-1$) is a special case of quantum lattice NLS model.

In the next sections, following the procedures in \cite{Korepin93} and \cite{Yang1968}, we construct thermodynamics and also find the CFT description of the model. In section 2, we describe zero temperature dynamics and elementary excitations. In section 3, we give CFT description and quenches of the model. In section 4, we discuss thermodynamics as well as zero temperature, and strong coupling cases. In section 5, we generalize our results to arbitrary negative spin. In section 6, we provide conclusions.

\section{The thermodynamic limit at zero temperature}
\setcounter{equation}{0}
Let us consider the holomorphic QCD model. In the thermodynamic limit, both the quantities $L$ and $N$ tend to infinity, while their ratio remains finite $D={N/L}$.

We first indicate that the following properties hold for the system: all the solutions $\lambda_j$ of Bethe equations (\ref{Beq}) are real numbers;
the solution of logarithmic form Bethe equations (\ref{Log-Bethe-eq}) exists and can be uniquely parameterized by a set of integer (half-integer) numbers.
Their proofs are given in Appendix A, and follow from \cite{Korepin93,Yang1968}.
The solution $\lambda_k$ is monotonically increasing with the corresponding number $n_k$. Specifying $\{n_k\}$ for the ground state (\ref{nj}), we obtain from (\ref{Log-Bethe-eq}):
\begin{equation}
2\pi\rho_t(\lambda)={1\over{L}}\sum^{N}_{j=1}K(\lambda(x),\lambda_j)+K(\lambda(x)).
\label{rho-t}\end{equation}
Here the density of vacancies is defined in Appendix B. It can be expressed as the sum of the densities of particles and holes
\begin{equation}
\rho_t(\lambda)=\rho_p(\lambda)+\rho_h(\lambda).
\end{equation}
The detailed analysis is given in Appendix B.

At zero temperature, the ground state with the lowest energy corresponds to the solutions $\lambda_j$ of the following Bethe equations, in which numbers $n_j$ are chosen from the equation (\ref{nj}),
\begin{equation}
L\;{\theta}(\lambda_j)+\sum^{N}_{l=1}\theta(\lambda_j-\lambda_l)=2\pi \left[j-\left(\frac{N+1}{2}\right)\right],\quad j=1,\cdots,N.
\label{Ground-state-BAE}\end{equation}
In the thermodynamic limit, the values of $\lambda_j$ are condensed in Fermi sphere, with the Fermi surface denoted by $q=\lim\lambda_N$. Then all the vacancies inside the interval $[-q,q]$ (called particles) are occupied by all the Bethe roots $\lambda_j$
(the density of holes $\rho_h(\lambda)=0$), see (\ref{rho-h-rho-p}).
One can get a linear integral equation for $\rho_p(\lambda)$ by changing the sum in (\ref{rho-t}) to an integral (see Appendix B)
\begin{equation}
2\pi\rho_p(\lambda)=\int^{q}_{-q}K(\lambda,\mu)\rho_p(\mu)d\mu+K(\lambda).
\label{rho-p}\end{equation}
In our case we have to require that
\begin{equation}
\int^q_{-q}\rho_p(\lambda)d\lambda=D={N\over{L}}.
\label{Density}\end{equation}
One can calculate the Fermi momentum as a unique function of $D$. Furthermore, one can also calculate the Fermi velocity $v_F$ also known as the quench velocity for local quenches. The calculation is based on the existence of Fermi surface $q$. Details will be shown in the next section.

Here we give some preliminary analysis of the dependence of $D$ on $q$.
The equations (\ref{rho-p}) and (\ref{Density}) show that the density $D$ is a function of the parameter $q$.
Obviously, when $q=0$, then $D=0$. On the other hand, when $q=\infty$, then $D>1$. This can be obtained by the following estimate.
Both two terms on the right hand side of equation (\ref{rho-p}) are positive.
Substituting them to equation (\ref{Density}) to calculate $D$ at $q=\infty$. Even the integral of second term will give rise to
\begin{equation}
{1\over2\pi}\int^\infty_{-\infty}K(\lambda)d\lambda=1,
\end{equation}
which means
\begin{equation}
\int^\infty_{-\infty}\rho_p(\lambda)d\lambda>1.
\end{equation}
Meanwhile, we show that $\partial D/\partial q>0$. The proof is the same as that of the Bose gas case in book \cite{Korepin93}. Briefly, by differentiating (\ref{Density}) with respect to $q$, one gets
\begin{equation}
\frac{\partial D}{\partial q}=\rho_p(q)+\rho_p(-q)+\int^q_{-q}\frac{\partial \rho_p(\lambda)}{\partial q}d\lambda>0.
\end{equation}
The equation (\ref{rho-p}) can also be differentiated to give an integral equation of $\partial \rho_p(\lambda)/\partial q$. Substituting the solution of integral equation into the above equation gives the explicit expression for $\partial D/\partial q$, and the inequality $\partial D/\partial q>0$.
Given the above, density $D$ is a monotonically increasing function of $q$ (when $q=0$, then $D=0$; and when $q=\infty$, then $D>1$).

Considering the grand canonical ensemble, one changes the Hamiltonian $\cal H$ to
\begin{equation}
{\cal H}_h={\cal H}-hQ=\sum^L_{j=1}H_{j,j+1}-h\sum^L_{j=1}z_j\partial_j,
\end{equation}
where $Q$ is the operator for the number of particles and $h$ is the chemical potential.
The commutativity $[{\cal H},Q]=0$ is ensured by the conservation law established in \cite{Lipatov94,Faddeev-Korchemsky1995}.
Now the eigenvalues of ${\cal H}_h$ are
\begin{equation}
E_h=\sum^N_{j=1}(\frac{-2}{\lambda^2_j+1}-h).
\end{equation}

In the framework of grand canonical ensemble we can consider elementary excitations\footnote{The book \cite{Korepin93} shows that for elementary excitation we have to change periodic boundary conditions to anti-periodic boundary conditions.}.
In order to achieve this, we define the dressed energy of elementary excitation $\varepsilon(\lambda)$ as the solution of the linear integral equation
\begin{equation}
\varepsilon(\lambda)-\frac{1}{2\pi}\int^{+q}_{-q}
K(\lambda,\mu)\varepsilon(\mu)d\mu=\frac{-2}{\lambda^2+1}-h\equiv\varepsilon_0(\lambda),
\label{linear-inte-eq}\end{equation}
with condition
\begin{equation}
\varepsilon(q)=\varepsilon(-q)=0,
\end{equation}
which defines the dependence of $q$ on $h$. Since this excitation emerges from the theory first proposed by Lipatov, we will call it ``lipaton". Formula (\ref{kp}) shows the relation of dressed (physical) momentum $k$ and bare momentum $p_0$.
The formulae (\ref{linear-inte-eq}) and (\ref{kp}) define bare energy $\varepsilon_0(\lambda)$ of lipaton. We can also write
\begin{equation}
\varepsilon_0(p_0)=-h-1-\cos(p_0).
\end{equation}

The density $D$ is also defined by $h$. This comes from the dependence of $\varepsilon(\lambda)$ on $h$.
The density is a monotonically increasing function of the chemical potential when $h>-2$
\begin{equation}
\frac{\partial D}{\partial h}>0,\quad D|_{h=-2}=0,\quad D|_{h=\infty}=\infty.
\end{equation}
For $h<-2$, the density is zero. We remark that, $\infty$ density corresponds to some positive limited value of $h$.
For details please see section $4$ and the book \cite{Korepin93}.

In section $4$, we shall consider the thermodynamics at nonzero temperature. The equation of $\varepsilon(\lambda)$ will be derived in a natural way in the zero temperature limit.

The function $\varepsilon(\lambda)$ is an even function of spectral parameter $\lambda$ monotonic on the positive semi-axis $\lambda>0$ and also
\begin{eqnarray}
&&\varepsilon(\lambda)>0,\quad\mbox{when}\quad |\lambda|>q;\\
&&\varepsilon(\lambda)<0,\quad\mbox{when}\quad |\lambda|<q.
\end{eqnarray}
When $h<-2$, the function $\varepsilon(\lambda)$ has no zeros on the real axis. This corresponds to the case of zero density, $D=0$ and $q=0$.
When $h>-2$, the function $\varepsilon(\lambda)$ has two zeros on the real axis
\begin{equation}
\varepsilon(\pm q)=0,\quad h>-2.
\end{equation}

Next subsection will show that $\varepsilon(\lambda)$ is the energy of the particle excitation above the ground state energy.
The meaning of $\varepsilon(\lambda)$ will be further clarified in section $4$.

\subsection{Excitations at zero temperature}

We already know that the ground state can be described by a set of integers $n_j$ given by (\ref{nj}), while the excited states are described by different sets of integers.
Let us consider the elementary particle excitation and hole excitation.
Due to the existence of excited particle $\lambda_p>|q|$ and hole $-q<\lambda_h<q$, the values of the solutions are changed $\lambda_j\rightarrow\tilde{\lambda}_j$.
The Bethe equations for the vacuum particles now become
\begin{equation}
L\;{\theta}(\tilde{\lambda}_j)+\sum_{l}\theta(\tilde{\lambda}_j-\tilde{\lambda}_l)
+\theta(\tilde{\lambda}_j-\lambda_p)-\theta(\tilde{\lambda}_j-\lambda_h)=2\pi \left[j-\left(\frac{{N}+1}{2}\right)\right],
\end{equation}
Subtracting this from the Bethe equations for the ground state (\ref{Ground-state-BAE}), and noticing that $\lambda_j-\tilde{\lambda}_j=O({L}^{-1})$ and $\theta(\lambda+\bigtriangleup)-\theta(\lambda)=O(\bigtriangleup)$, ($\bigtriangleup$ here is not $\Delta$) one has,
\begin{eqnarray}
&&L(\lambda_j-\tilde{\lambda}_j)K(\lambda_j)-\theta({\lambda}_j-{\lambda}_p)
+\theta({\lambda}_j-{\lambda}_h)\nonumber\\
&&~~~~+(\lambda_j-\tilde{\lambda}_j)\sum_{l}K({\lambda}_j,{\lambda}_l)
-\sum_{l}K({\lambda}_j,{\lambda}_l)(\lambda_l-\tilde{\lambda}_l)=0.
\end{eqnarray}
By using equation (\ref{rho-p}), one obtains
\begin{eqnarray}
&&2\pi\rho_p(\lambda_j)L(\lambda_j-\tilde{\lambda}_j)-\theta({\lambda}_j-{\lambda}_p)
+\theta({\lambda}_j-{\lambda}_h)\nonumber\\
&&~~~~-\sum_{l}K({\lambda}_j,{\lambda}_l)(\lambda_l-\tilde{\lambda}_l)
\frac{(\lambda_{l+1}-\lambda_l)}{(\lambda_{l+1}-\lambda_l)}=0.
\label{shifted-BAE}\end{eqnarray}
Let us introduce the shift function $F$ for one particle and one hole:
\begin{equation}
F(\lambda_j|\lambda_p,\lambda_h)\equiv
\frac{(\lambda_j-\tilde{\lambda}_j)}{(\lambda_{j+1}-\lambda_j)}.
\end{equation}
Let us discuss the meaning of $F$ function. When producing a hole, all other $\lambda_j$ will shift a little bit.
$F$ function describes these shifts divided by the normal distances in the ground state.
Thus $F$ function is a reaction function which shows the reaction of the Fermi sphere (distribution of particles) after the creation of a hole.

In the thermodynamic limit, one can change the sum in (\ref{shifted-BAE}) to integral, which gives
\begin{equation}
F(\mu|\lambda_p,\lambda_h)-
\int^q_{-q}\frac{d\nu}{2\pi}K(\mu,\nu)F(\nu|\lambda_p,\lambda_h)=
\frac{\theta(\mu-\lambda_p)-\theta(\mu-\lambda_h)}{2\pi}.
\end{equation}
Physical meaning of the shift function becomes clear, when considering scattering (see the end of the section, equations (\ref{S}) and (\ref{phi})). It is simply related to the scattering phase:
\begin{equation}
F(\mu|\lambda_p,\lambda_h) =
\frac{\phi(\mu, \lambda_p)-\phi(\mu, \lambda_h)}{2\pi}.
\end{equation}
Now the dressed energy also called the one particle excitation energy (elementary excited) $\varepsilon(\lambda)$ is
\begin{equation}
\varepsilon(\lambda)-\frac{1}{2\pi}\int^{+q}_{-q}
K(\lambda,\mu)\varepsilon(\mu)d\mu=\frac{-2}{\lambda^2+1}-h\equiv\varepsilon_0(\lambda),
\label{dressed-E}\end{equation}
with condition
\begin{equation}
\varepsilon(q)=\varepsilon(-q)=0.
\end{equation}
The observable energy $\delta E$ is equal to the energy of the excited state minus the ground state:
\begin{eqnarray}
\delta E(\lambda_p,\lambda_h)&=&\varepsilon_0(\lambda_p)-\varepsilon_0(\lambda_h)
+\sum_j[\varepsilon_0(\tilde{\lambda}_j)-\varepsilon_0(\lambda_j)]\nonumber\\
&=&\varepsilon_0(\lambda_p)-\varepsilon_0(\lambda_h)
-\int^q_{-q}\varepsilon'_0(\mu)F(\mu|{\lambda}_p,\lambda_h)d\mu\nonumber\\
&=&\varepsilon(\lambda_p)-\varepsilon(\lambda_h).
\end{eqnarray}
The proof for the last step in above equation is the same as that in Chapter I Section 4 of the book \cite{Korepin93}.

The momentum of the particle $k(\lambda_p)$ is
\begin{eqnarray}
k(\lambda_p)=p_0(\lambda_p)+\int^q_{-q}\theta(\lambda_p-\mu)\rho_p(\mu)d\mu,\qquad
p_0(\lambda)=i\ln\left(\frac{i+\lambda}{i-\lambda}\right),
\label{kp}\end{eqnarray}
where $p_0(\lambda)$ is the bare momentum\footnote{Note that $p_0(\lambda)$ is an odd function of $p_0(-\lambda)=-p_0(\lambda)$.}.
The value $\lambda_p$ of elementary particle excitation is out of Fermi sphere, $|\lambda_p|>q$.
On the contrary, the momentum $k(\lambda_h)$ of elementary hole excitation with energy $-\varepsilon(\lambda_h)$ is
\begin{eqnarray}
k_h(\lambda_h)=-p_0(\lambda_h)-\int^q_{-q}\theta(\lambda_h-\mu)\rho_p(\mu)d\mu.
\label{ph}\end{eqnarray}
where $-q<\lambda_h<q$.
\begin{figure}[H]
\centering
\includegraphics[width=0.5\textwidth]{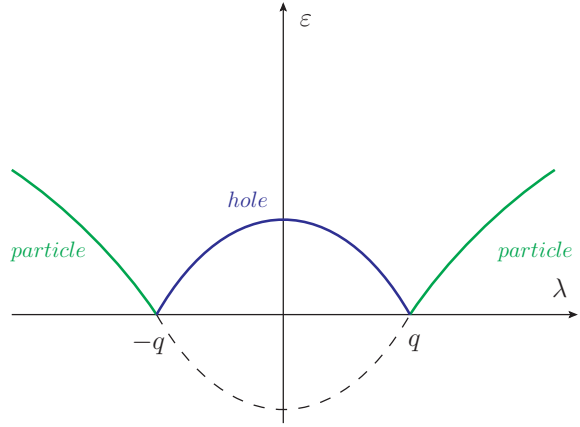}
\caption{The energy of the lipaton as a function of $\lambda$. \protect\\For $-q<\lambda<q$ lipaton is a hole, but it is the particle for other values of $\lambda$.}
\label{fig1}
\end{figure}
Similarly, the observable momentum ${\delta}P(\lambda_p,\lambda_h)$ is
\begin{eqnarray}
{\delta}P(\lambda_p,\lambda_h)&=&p_0(\lambda_p)-p_0(\lambda_h)-\int^q_{-q}F(\mu|{\lambda}_p,\lambda_h)d\mu\nonumber\\
&=&p_0(\lambda_p)-p_0(\lambda_h)
+\int^q_{-q}[\theta(\lambda_p-\mu)-\theta(\lambda_h-\mu)]\rho_p(\mu)d\mu.
\end{eqnarray}
Since  particle and hole are defined at different momenta, we can combine them together in one elementary excitation which we will call lipaton, see Figure \ref{fig1}.
It is similar to continuous nonlinear Schroedinger's equation, considered in the book   \cite{Korepin93}.
Following this approach we can prove that in the sector with fixed density any energy level of lattice  nonlinear Schroedinger's equation \footnote{Equivalent to XXX with negative spin.} is a scattering state of several elementary excitations [lipatons], see formula (4.29) in Chapter I section 4 of the book \cite{Korepin93}.
The lipaton is a fermion. It can be represented as a topological excitation (soliton) of original bosonic degrees of freedom, described by the group $Z_2$ . Topological solitons were described in   \cite{Faddeev}.
In quantum nonlinear Schroedinger's equation topological excitation was first described  in 1993, see  Chapter I section 4 of the book \cite{Korepin93}.

The LNS model has infinitely many conservation laws. This impose strong limitations on scattering.
For example there is no reflection during scattering of particle and a hole.
The \textbf{scattering matrix} is a transition coefficient:
\begin{eqnarray}
S=\exp\{-i{\phi}(\lambda_p,\lambda_h)\},
\label{S}\end{eqnarray}
here the scattering phase is defined by  the following integral equation:
\begin{eqnarray}
\phi(\lambda_p,\lambda_h)-\frac{1}{2\pi}\int^{+q}_{-q}
K(\lambda_p,\nu)\phi(\nu,\lambda-\lambda_h)d\nu=\theta(\lambda_p-\lambda_h).
\label{phi}\end{eqnarray}
Many body scattering matrix is a product of two-body scattering matrices.

\section{Finite size effects and CFT}
\setcounter{equation}{0}
In this section, we show that the critical exponent $\vartheta$
which drives the long distance asymptotics of correlation functions can be expressed in terms of dressed charge $Z(\lambda)$. We further discuss that the model under consideration can be described by two conformal dimensions $\Delta^{\pm}$, and they are related to two Virasoro algebras both with central charge equals to $1$.
\begin{figure}[H]
\centering
\includegraphics[width=0.5\textwidth]{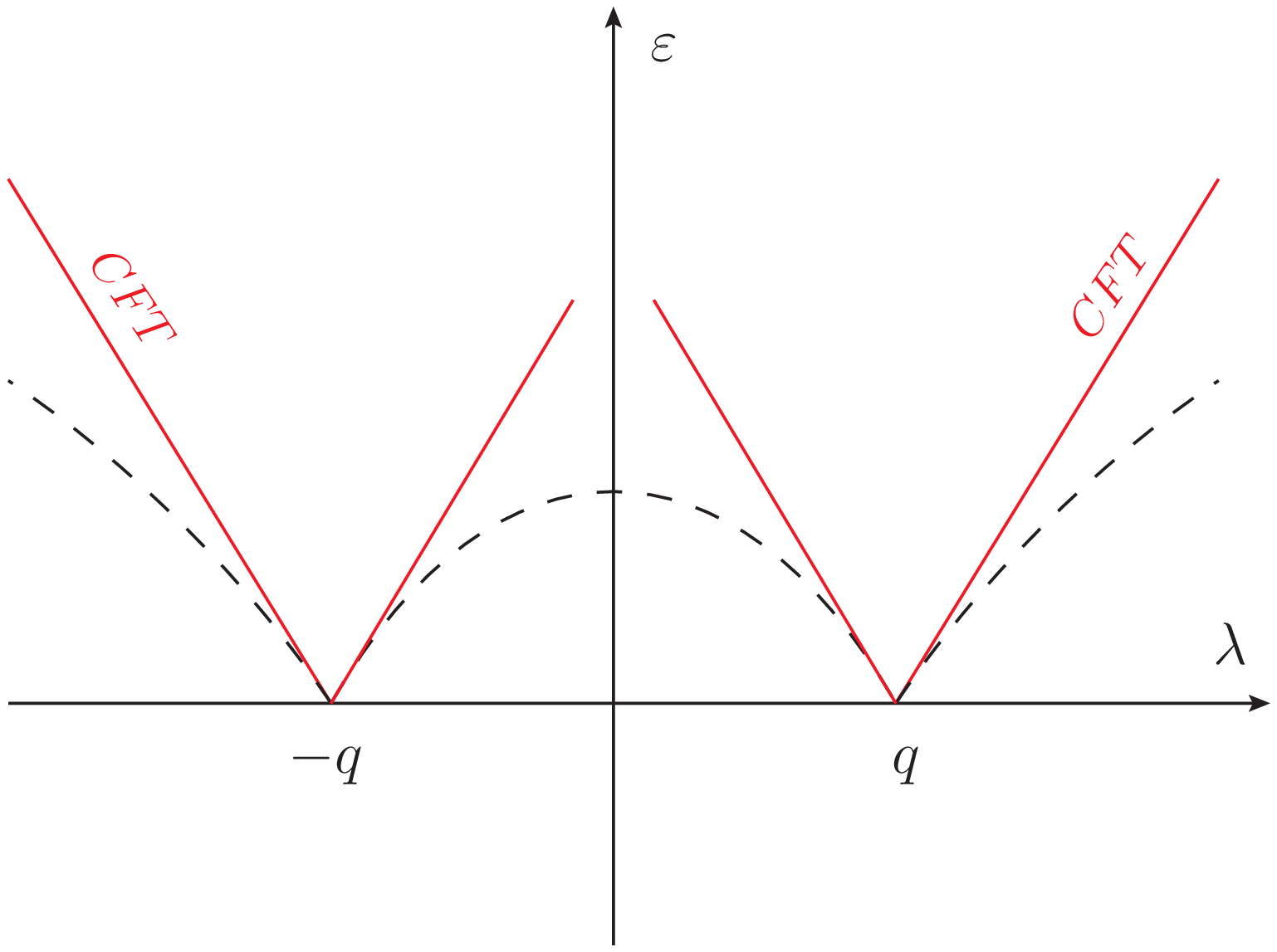}
\caption{CFT approximation to $\varepsilon$. \protect\\The slope of the straight lines is $\partial\varepsilon(\lambda)/\partial\lambda|_{\lambda=q}$. It is the first factor of the equation (\ref{vf}) for Fermi velocity.}
\label{fig2}\end{figure}

In the thermodynamic limit, the elementary excitation energy satisfies the integral equation
\begin{eqnarray}
\varepsilon(\lambda)=\frac{-2}{\lambda^2+1}-h+\frac{1}{2\pi}\int^{q}_{-q}
K(\lambda,\mu)\varepsilon(\mu)d\mu,
\label{dressed-energy}\end{eqnarray}
and the energy of the hole (on the Fermi sphere) should be equal to
zero, determining the Fermi spectral parameter $q$
\begin{equation}
\varepsilon(\pm q)=0.
\end{equation}
The fractional charge $Z(\lambda)$ is defined by the dressing equation
\begin{equation}
Z(\lambda)-{1\over2\pi}\int^q_{-q}K(\lambda,\mu)Z(\mu)d\mu=1.
\label{Z}\end{equation}
The physical meaning of the fractional charge comes from differentiating of linear integral equation of $\varepsilon(\lambda)$ with respect to $h$,
\begin{equation}
Z(\lambda)=-\frac{\partial\varepsilon(\lambda)}{\partial h}.
\end{equation}
The ground state is the Fermi sphere filled with the following states: $-q\leq\lambda\leq q$.

The spectrum of conformal dimensions $\Delta^{+}$ and $\Delta^{-}$ of the primary field are given by
\begin{equation}
2\Delta^{\pm}=2N^{\pm}+\left(\frac{\delta{N}}{2\cal{Z}}\pm{\cal{Z}}d\right)^2,
\end{equation}
where $\cal{Z}$ is the value of fractional charge $Z(\lambda)$ on the Fermi surface $q$
\begin{equation}
{\cal Z}=Z(q).
\label{calZ}\end{equation}
$N^{\pm}$ show the change of integer numbers $n_j$ (periodic boundary conditions of the Bethe Ansatz) in the vicinity of the Fermi surface $\pm q$.
The quantum number $\delta{N}$ is characteristic of the local conformal field.
The quantum number $d$ represents the number of backscattered particles .

We remark that the asymptotics of correlation functions at zero temperature can be expressed in terms of conformal dimensions.
It should also be mentioned that since the dressed charge $\cal{Z}$ depends only on the R-matrix (and Fermi momentum), the conformal dimensions demonstrate universality properties.

The Fermi velocity $v_F$ (also known as quench velocity $v_e$ for local quenches \cite{Essler2014,Bose2018}) can be defined as
\begin{equation}
v_F=v_e=\frac{\partial\varepsilon(\lambda)}{\partial k(\lambda)}\bigg|_{\lambda=q}=\frac{\partial\varepsilon(\lambda)}{\partial\lambda}\bigg|_{\lambda=q}
\left(\frac{\partial k(\lambda)}{\partial\lambda}\right)^{-1}\bigg|_{\lambda=q},
\label{vf}\end{equation}
which strongly depends on the Fermi surface\footnote{which depends on the density (\ref{Density})} $q$, see Figure \ref{fig2}. We have discussed the value range of $q$ and its influence in the last section.
Here $\varepsilon(\lambda)$ is the dressed energy (\ref{dressed-energy}) and $k(\lambda)$ is the
dressed momentum of the excitation:
\begin{eqnarray}
k(\lambda)=p_0(\lambda)+\int^q_{-q}\theta(\lambda-\mu)\rho_p(\mu)d\mu,\qquad
p_0(\lambda)=i\ln\left(\frac{i+\lambda}{i-\lambda}\right).
\end{eqnarray}
So
\begin{equation}
\frac{\partial k(\lambda)}{\partial\lambda}=K(\lambda)+\int^q_{-q}K(\lambda,\mu)\rho_p(\mu)d\mu
=2\pi\rho_t(\lambda)=2\pi \rho_p (\lambda).
\end{equation}
Thus $v_F$ can be represented in the following form:
\begin{equation}
v_F=\frac{1}{2\pi\rho_t(q)}
\frac{\partial\varepsilon(\lambda)}{\partial\lambda}\bigg|_{\lambda=q}
=\frac{1}{2\pi\rho_p(q)}
\frac{\partial\varepsilon(\lambda)}{\partial\lambda}\bigg|_{\lambda=q};
\end{equation}
note that the density of holes at Fermi surface $q$ should be equal to zero.

In the Euclidian formulation, the correlation functions depend on $|x+iv_{F} t|$, where $x$ and $t$ are space and time, see Chapter XVIII Section 2 of the book \cite{Korepin93}. The {Fermi velocity $v_F$ coincides with $v_e$, the velocity of the spread of entanglement entropy after a measurement quench}, see \cite{Bose2018}.

We can define the critical exponent in the case of XXX spin $s=-1$ chain,
\begin{equation}
\vartheta=2Z^2(q)=2{\cal Z}^2,
\end{equation}
see (\ref{Z}) and (\ref{calZ}). This function is important because it defines the power law decrease of the long distance asymptotics of the correlation functions at zero temperature (see \cite{Korepin93,Izergin-Korepin-Reshetikhin-1989,Bogoliubov1987}).
The results of this section can thus be applied to the calculation of the asymptotics of correlation functions.

At zero temperature, the entanglement entropy \cite{Korepin2004} of the model is
\begin{equation}
{\cal S}(y)={1\over3}\ln y,\;\text{as}\;y\rightarrow\infty,
\end{equation}
where $y$ is the size of subsystem; for current model the central charge is equal to $1$. This result agrees with the third law of thermodynamics.

The critical exponents in a conformally invariant theory are related to the
scaling (conformal) dimensions of the operators within this theory. Bethe
Ansatz solvable models (which we consider here) are related to two Virasoro
algebras, both with central charge equals to $1$.

\section{Thermodynamics}
\setcounter{equation}{0}

In the following, we mainly focus on thermodynamics of XXX spin chain with spin $s=-1$, and then generalize the results to arbitrary negative spin $s$ in the next section.
The Bethe equations (and roots) are of primary importance. Their properties are discussed in section 2 and Appendix A and B. They are the basis of this section.

Following the works \cite{Korepin93,Yang1968}, one derives the Yang-Yang equation for current model by using variation and the steepest descent method,
\begin{equation}
\varepsilon(\lambda)=\frac{-2}{\lambda^2+1}-h-\frac{T}{2\pi}\int^{+\infty}_{-\infty}
K(\lambda,\mu)\ln(1+e^{-{\varepsilon(\mu)/ T}})d\mu.
\end{equation}
The existence of solution for this equation is given in Appendix C.
Simultaneously, the following properties are also established:
\begin{eqnarray}
&&\frac{\rho_h(\lambda)}{\rho_p(\lambda)}=e^{\varepsilon(\lambda)/ T};\\
&&D={N\over L}=\int^\infty_{-\infty}\rho_p(\lambda)d\lambda;\\
&&\varepsilon(\lambda)=\varepsilon(-\lambda);\\
&&\frac{-2}{\lambda^2+1}-h\geq\varepsilon(\lambda)\geq\frac{-2}{\lambda^2+1}+2+x_0;\label{estimate}\\
&&\frac{\partial\varepsilon(\lambda)}{\partial\lambda}>0,\quad \text{if}\,\lambda>0;\\
&&\varepsilon(\lambda)\rightarrow\frac{-2}{\lambda^2+1}-h+O(1/\lambda^2),\quad\lambda\rightarrow\pm\infty.
\end{eqnarray}

Based on Yang-Yang equation, one can further derive some physical quantities like free energy, pressure and entropy. The free energy density is
\begin{eqnarray}
f=\int^{+\infty}_{-\infty}(\frac{-2}{\mu^2+1}-\varepsilon(\mu))\rho_p(\mu)
-T(\rho_p(\mu)+\rho_h(\mu))\ln(1+\exp(-{\varepsilon(\mu)/{T}}))d\mu.
\end{eqnarray}
Substitute Yang-Yang equation into it (in the following equation all integrals are from $-\infty$ to $+\infty$):
\begin{eqnarray}
f&=&\int\left[h+\frac{T}{2\pi}\int\frac{2\kappa\,d\nu}{\kappa^2+(\mu-\nu)^2}\ln(1+\exp(-{\varepsilon(\nu)/{T}}))\right]\rho_p(\mu)d\mu\nonumber\\
&&-T\int(\rho_p(\mu)+\rho_h(\mu))\ln(1+\exp(-{\varepsilon(\mu)/{T}}))d\mu\nonumber\\
&=&\int\left[h\rho_p(\mu)-T(\rho_p(\mu)+\rho_h(\mu))\ln(1+\exp(-{\varepsilon(\mu)/{T}}))\right]d\mu\nonumber\\
&&+\frac{T}{2\pi}\int\left[\frac{2\kappa\rho_p(\nu)\,d\nu}{\kappa^2+(\mu-\nu)^2}\right]\ln(1+\exp(-{\varepsilon(\mu)/{T}}))d\mu\nonumber\\
&=&\int{h}\rho_p(\mu)d\mu-T\ln(1+\exp(-{\varepsilon(\mu)/{T}}))\nonumber\\
&&\times\left[(\rho_p(\mu)+\rho_h(\mu))-\frac{1}{2\pi}\int\frac{2\kappa\rho_p(\nu)\,d\nu}{\kappa^2+(\mu-\nu)^2}\right]d\mu.
\end{eqnarray}
Then substitute the expression for $\rho_t(\lambda)$ into the above equation
\begin{equation}
\rho_t(\lambda)=\rho_p(\lambda)+\rho_h(\lambda)={1\over2\pi}\left[\int^{+\infty}_{-\infty}K(\lambda,\nu)\rho_p(\nu)d\nu+K(\lambda)\right].
\end{equation}
One gets the simplified expression for free energy density
\begin{eqnarray}
f&=&\int^{+\infty}_{-\infty}\left[h\rho_p(\mu)-\frac{T}{2\pi}\ln(1+\exp(-{\varepsilon(\mu)/{T}}))K(\mu)\right]d\mu\nonumber\\
&=&D h-\frac{T}{2\pi}\int^{+\infty}_{-\infty}\ln(1+\exp(-{\varepsilon(\mu)/{T}}))K(\mu)d\mu,
\end{eqnarray}
in which $\int\rho_p(\mu)d\mu=D={N/L}$. Then the free energy becomes:
\begin{equation}
{\cal F}=Nh-\frac{LT}{2\pi}\int^{+\infty}_{-\infty}\ln(1+\exp(-{\varepsilon(\mu)/{T}}))K(\mu)d\mu.
\end{equation}
The pressure is the derivative of the free energy with respect to $L$ at fixed temperature:
\begin{equation}
{\cal P}=-\left(\frac{\partial {\cal F}}{\partial L}\right)_T=\frac{T}{2\pi}\int^{+\infty}_{-\infty}
K(\mu)\ln(1+e^{-{\varepsilon(\mu)/ T}})d\mu.
\end{equation}
The total entropy of the whole system can be expressed as:
\begin{equation}
S=-\frac{\partial {\cal F}}{\partial T}=\frac{L}{2\pi}\int^{+\infty}_{-\infty}
K(\mu)\left[\ln(1+e^{-{\varepsilon(\mu)/ T}})+\frac{\varepsilon(\mu)}{T(e^{\varepsilon(\mu)/ T}+1)}\right]d\mu.
\end{equation}
\subsection{Zero temperature $T\rightarrow0$ limit}
Let us first discuss  what happens with the estimate (\ref{estimate}) to the solution of the Yang-Yang equation $\varepsilon(\lambda)$ under the zero temperature $T\rightarrow0$ limit.
According to (\ref{h}), the constant $x_0$ is a one to one monotonically decreasing function of $h$. When $T=0$, the point $h=-2$ corresponds to $x_0=0$. The semi-axis are mapped in the following way:
\begin{eqnarray}
&&\mbox{when}\quad h>-2, \quad\mbox{then}\quad x_0<0;\\
&&\mbox{when}\quad h<-2, \quad\mbox{then}\quad x_0>0.
\end{eqnarray}
In the last case, it is obvious that $h=-2-x_0$. The function $\varepsilon(\lambda)$ is then equal to
\begin{equation}
\varepsilon(\lambda)=\frac{-2}{\lambda^2+1}-h,\quad h<-2.
\end{equation}
The function has no zeros on the real axis. This corresponds to the case of zero density, $D=0$.
When $h>-2$, then $x_0<0$, and this is the case that $\varepsilon(\lambda)$ possesses two zeros on the real axis
\begin{equation}
\varepsilon(\pm q)=0,\quad 0>h>-2.
\end{equation}
The function $\varepsilon(\lambda)$ is an even function of spectral parameter $\lambda$ monotonic on the positive semi-axis $\lambda>0$ and also
\begin{eqnarray}
&&\mbox{when}\quad |\lambda|>q,\quad\mbox{then}\quad\varepsilon(\lambda)>0;\\
&&\mbox{when}\quad |\lambda|<q,\quad\mbox{then}\quad\varepsilon(\lambda)<0.
\end{eqnarray}

These properties are obtained from in the limit $T\rightarrow0$. The Yang-Yang equation becomes linear in this case

\begin{eqnarray}
&&\varepsilon(\lambda)=\frac{-2}{\lambda^2+1}-h+\frac{1}{2\pi}\int^{+q}_{-q}
K(\lambda,\mu)\varepsilon(\mu)d\mu\\
&&\varepsilon(q)=\varepsilon(-q)=0.
\end{eqnarray}

During this procedure, we have used the fact that
\begin{eqnarray}
e^{\varepsilon(\lambda)/T}=\frac{\rho_h(\lambda)}{\rho_p(\lambda)}=
\left\{
  \begin{array}{cc}
    \infty, & |\lambda|>q \\
    0, & |\lambda|<q \\
  \end{array}
\right.\nonumber\\
\text{if }-q\leq\lambda\leq q, \text{ then }\rho_h(\lambda)=0 \text{ and } \rho_p(\lambda)>0\label{rho-h-rho-p}\\
\text{if }\lambda<-q \text{ or } \lambda>q, \text{ then } \rho_p(\lambda)=0 \text{ and } \rho_h(\lambda)>0.\nonumber
\end{eqnarray}
The equation for densities becomes
\begin{eqnarray}
&&2\pi\rho_p(\lambda)-\int^q_{-q}K(\lambda,\mu)\rho_p(\mu)d\mu=K(\lambda);\quad -q\leq\lambda\leq q\\
&&\int^q_{-q}\rho_p(\lambda)d\lambda=D={N\over{L}}.
\end{eqnarray}
At zero temperature, the free energy is:
\begin{equation}
{\cal F}=Nh+\frac{L}{2\pi}\int^q_{-q}\frac{2\kappa}{\kappa^2+\mu^2}\varepsilon(\mu)d\mu.
\end{equation}
The pressure ${\cal P}$ is equal to
\begin{eqnarray}
{\cal P}=-\frac{1}{2\pi}\int^q_{-q}\frac{2\kappa}{\kappa^2+\mu^2}\varepsilon(\mu)d\mu.
\end{eqnarray}
The entropy vanishes (note that $\varepsilon(\lambda)<0$ in the interval $|\lambda|<q$, so $\varepsilon(\mu)/T\rightarrow-\infty$ as $T\rightarrow0$):
\begin{equation}
S=-\frac{\partial {\cal F}}{\partial T}=\frac{L}{2\pi}\int^{+q}_{-q}
\frac{2\kappa}{\kappa^2+\mu^2}\left[-{\varepsilon(\mu)\over T}+\frac{\varepsilon(\mu)}{T(e^{\varepsilon(\mu)/T}+1)}\right]d\mu=0,
\end{equation}
according to the third law of thermodynamics.

\subsection{Strong coupling for lattice nonlinear Schroedinger}
In another limiting case, consider temperature  $T$ to be  arbitrary, but coupling is strong ($\kappa\rightarrow\infty$). As $\kappa\rightarrow\infty$, the kernel $K(\lambda,\mu)$ tends to zero and all the integral equations can be simplified:
\begin{eqnarray}
&&\varepsilon(\lambda)=\frac{-2}{\lambda^2+1}-h-{\cal P}+O\left({1\over\kappa^3}\right);\\
&&\rho_p(\lambda)=\frac{1+D}{\kappa\pi(1+e^{\varepsilon(\lambda)/T})};\\
&&\rho_h(\lambda)=\frac{1+D}{\kappa\pi(1+e^{-\varepsilon(\lambda)/T})};\\
&&\rho_t(\lambda)=\frac{1+D}{\kappa\pi};\\
&&D={N\over{L}}=\int^\infty_{-\infty}\rho_p(\lambda)d\lambda.
\end{eqnarray}

The elementary excitation [lipaton] is a fermion. Its momentum $k$ and energy $\varepsilon$ depend on spectral parameter $\lambda$. We can exclude $\lambda$ and write \footnote{in the strong coupling limit} energy of lipaton as a function of momentum
\begin{equation}
\varepsilon(k)=-h-1-\cos(k).
\end{equation}

\section{Generalized negative spin $s=-|s|$ case}
\setcounter{equation}{0}
In order to fully identify XXX with NLS we have to consider general negative spin.
For arbitrary negative spin $s=-|s|$, based on the logarithmic form Bethe equations (\ref{Log-Bethe-eq-NLS}), the corresponding Yang-Yang equation becomes
\begin{equation}
\varepsilon(\lambda)=\frac{-2|s|}{\lambda^2+s^2}-h-\frac{T}{2\pi}\int^{+\infty}_{-\infty}
K(\lambda,\mu)\ln(1+e^{-{\varepsilon(\mu)/ T}})d\mu.
\end{equation}
The free energy is
\begin{equation}
{\cal F}=Nh-\frac{TL}{2\pi}\int^{+\infty}_{-\infty}
\frac{2|s|\kappa}{(s\kappa)^2+\mu^2}\ln(1+e^{-{\varepsilon(\mu)/ T}})d\mu.
\end{equation}
The pressure is the derivative of the free energy with respect to $L$ at fixed temperature:
\begin{equation}
{\cal P}=-\left(\frac{\partial {\cal F}}{\partial L}\right)_T=\frac{T}{2\pi}\int^{+\infty}_{-\infty}
\frac{2|s|\kappa}{(s\kappa)^2+\mu^2}\ln(1+e^{-{\varepsilon(\mu)/ T}})d\mu.
\end{equation}
The entropy can be expressed as:
\begin{equation}
S=-\frac{\partial {\cal F}}{\partial T}=\frac{L}{2\pi}\int^{+\infty}_{-\infty}
\frac{2|s|\kappa}{(s\kappa)^2+\mu^2}\left[\ln(1+e^{-{\varepsilon(\mu)/ T}})+\frac{\varepsilon(\mu)}{T(e^{\varepsilon(\mu)/T}+1)}\right]d\mu.
\end{equation}
Further discussions and analysis for spin $s=-1$ case can be applied to arbitrary negative spin case, the procedure is similar, so they are omitted here.


\section{Conclusion}

The holomorphic QCD  describing the deep-inelastic scattering at small Bjorken $x$ is dual to the XXX spin $s=-1$ chain. It  is solvable by Bethe Ansatz. In this paper, we have constructed thermodynamics of this model and have generalized the results to arbitrary negative spin.
We also gave the description of the model in terms of CFT, and described the velocity, with which the entanglement entropy spreads after local quenches.
In thermodynamic limit any energy level is a scattering state of several elementary excitations\footnote{with different momenta},
see formula (4.29) in Chapter I section 4 of the book \cite{Korepin93}.
Here we call the elementary excitation lipaton.
The lipaton  is a fermion: it can be considered as a topological excitation of original [bosonic] degrees of freedom, described by  $Z_2$ group. Topological solitons were studied in the paper \cite{Faddeev}.
Topological excitation in nonlinear Schoedinger was  discovered   in \cite{Korepin93}.\par
The topological expansion in QCD implies the cylindrical topology of Feynman diagrams describing  high energy scattering; this results in the periodic boundary condition in the spin chain \cite{Gorsky02}. The lipaton satisfies anti-periodic boundary conditions: so only even number of lipatons will fit into periodic boundary conditions.
 We also calculated energy and momentum of lipaton, see (\ref{dressed-E}),  (\ref{kp}) and (\ref{ph}).
We have also calculated the scattering matrix of two  lipatons. \par\par
 In the forthcoming paper, we will address the implications of our results for the physics of deep-inelastic scattering, and for its modeling in quantum simulations.
\section*{Acknowledgments}

K.H. was supported by the National Natural Science Foundation
of China (Grant No. 11805152) and Shaanxi Key Laboratory for Theoretical Physics Frontiers in China.
The work of D.K. was supported by the U.S. Department of Energy, Office of Nuclear Physics, under contracts DE-FG-88ER40388 and DE-AC02-98CH10886, and by the Office of Basic Energy Science under contract DE-SC-0017662.
V.K. was supported by SUNY center for QIS at Long Island project number CSP181035.
The authors would like to thank Profs. Fabian Essler, Andrea Trombettoni, Abolfazl Bayat and Ctirad Klimchik for the helpful discussions.

\section*{Appendix A: Proofs of properties of the solutions of Bethe equations}
\setcounter{equation}{0}
\renewcommand{\theequation}{A.\arabic{equation}}

\textbf{Theorem 1.} All the solutions of Bethe equations (\ref{Beq}) for $s=-1$ are real numbers.
\begin{eqnarray}
\left({\frac{\lambda_k-i}{\lambda_k+i}}\right)^L
=\prod^{N}_{j=1,\ j\neq k} \frac{\lambda_k-\lambda_j+i}{\lambda_k-\lambda_j-i},\qquad
k=1,\cdots,N.
\end{eqnarray}
Proof: Both sides of the Bethe equations have the following properties.
\begin{eqnarray}
\text{LHS:} \left|\frac{\lambda-i}{\lambda+i}\right|\leq1,\quad\text{when Im} \lambda\geq0;\quad
\left|\frac{\lambda-i}{\lambda+i}\right|\geq1,\quad\text{when Im} \lambda\leq0;\label{estimate-lhs}\\
\text{RHS:} \left|\frac{\lambda+i}{\lambda-i}\right|\geq1,\quad\text{when Im} \lambda\geq0;\quad
\left|\frac{\lambda+i}{\lambda-i}\right|\leq1,\quad\text{when Im} \lambda\leq0;\label{estimate-rhs}
\end{eqnarray}
Consider the set of complex solutions $\{\lambda_j\}$ which satisfy the Bethe equations.
If denote the one with maximal imaginary part as $\lambda_{max}\in\{\lambda_j\}$, then
\begin{equation}
\text{Im }\lambda_{max}\geq \text{Im }\lambda_j,\quad j=1,\cdots,N.
\label{im-up-bound}\end{equation}
Taking the modulus of both sides of the equation for $\lambda_j=\lambda_{max}$, make use of the estimate for the right hand side (\ref{estimate-rhs}) and obtain
\begin{eqnarray}
\left|{\frac{\lambda_{max}-i}{\lambda_{max}+i}}\right|^L
=\left|\prod_{j=1}^N \frac{\lambda_{max}-\lambda_j+i}{\lambda_{max}-\lambda_j-i}\right|\geq1.
\end{eqnarray}
Due to (\ref{estimate-lhs}) and (\ref{im-up-bound}), this results in
\begin{equation}
\text{Im }\lambda_j\leq\text{Im }\lambda_{max}\leq0,\quad j=1,\cdots,N.
\end{equation}
On the other hand, let us define $\lambda_{min}$ as the one with minimal imaginary part: Im $\lambda_{min}\leq$ Im $\lambda_j$, $j=1,\cdots,N$. From the estimate for the right hand side (\ref{estimate-rhs}), we also derive
\begin{eqnarray}
\left|{\frac{\lambda_{min}-i}{\lambda_{min}+i}}\right|^L
=\left|\prod_{j=1}^N \frac{\lambda_{min}-\lambda_j+i}{\lambda_{min}-\lambda_j-i}\right|\leq1.
\end{eqnarray}
Similarly, due to (\ref{estimate-lhs}), one has
\begin{equation}
0\leq\text{Im }\lambda_{min}\leq\text{Im }\lambda_{j},\quad j=1,\cdots,N.
\end{equation}
Thus the only remaining possibility is Im $\lambda_j=0$, $j=1,\cdots,N$.
Theorem 1 is proved.
\\\\
\textbf{Theorem 2.} The solutions of the logarithmic form Bethe equations (\ref{Log-Bethe-eq}) exist.
\\
Proof: The proof procedure follows from \cite{Korepin93}, and it is based on the fact that the equations (\ref{Log-Bethe-eq}) can be obtained from a variational principle.
The Yang-Yang action was introduced by C.N. Yang and C.P. Yang in \cite{Yang1968}:
\begin{equation}
S=L\;\sum^N_{k=1}\theta_1(\lambda_k)+{1\over2}\sum^N_{k,j}\theta_1(\lambda_k-\lambda_j)
-2\pi\sum^N_{k=1}n_k\lambda_k,
\end{equation}
where $\theta_1(\lambda)=\int^\lambda_0\theta(\mu)d\mu$. Equations (\ref{Log-Bethe-eq}) are the extremum (minimum) conditions for $S$ $(\partial S/\partial\lambda_j=0)$.
To prove this, it is sufficient to establish that the matrix of second derivatives is positive definite.
\begin{equation}
\frac{\partial^2S}{\partial\lambda_j\partial\lambda_l}=\delta_{jl}[L\; K(\lambda_j)+\sum^N_{m=1}K(\lambda_j,\lambda_m)]-K(\lambda_j,\lambda_l)
\label{2nd-S-matrix}\end{equation}
where
\begin{equation}
K(\lambda,\mu)=\theta'(\lambda-\mu)=\frac{2\kappa}{\kappa^2+(\lambda-\mu)^2},\;\;K(\lambda)=K(\lambda,0),\quad \kappa=1,
\end{equation}
and so, one has
\begin{equation}
\sum_{j,l}\frac{\partial^2S}{\partial\lambda_j\partial\lambda_l}v_jv_l=
\sum^N_{j=1}L\; K(\lambda_j)v_j^2+\sum^N_{j>l}K(\lambda_j,\lambda_l)(v_j-v_l)^2\geq\sum^N_{j=1}L\; K(\lambda_j)v_j^2>0,
\end{equation}
for any vector $v_j$ with real components. The matrix of second derivatives is positive, the action $S$ is convex. $S$ has only one minimal extremum, and it defines the solutions of the Bethe equations. Theorem 2 is proved.

We remark that the determinant of the matrix (\ref{2nd-S-matrix}) gives the square of the norm of the Bethe wave function. The corresponding formula in the book \cite{Korepin93} is (0.3) in the Introduction section of Chapter X. So the wave function does note vanish.

\section*{Appendix B: Analysis of densities of momenta (the solutions of Bethe equations)}
\setcounter{equation}{0}
\renewcommand{\theequation}{B.\arabic{equation}}
To construct the thermodynamics, we shall need the following properties.
Let us subtract the $k-$th equation from $j-$th Bethe equation (\ref{Log-Bethe-eq}) of the system,
\begin{equation}
2\pi(n_j-n_k)=\sum^N_{l=1}[\theta(\lambda_j-\lambda_l)-\theta(\lambda_k-\lambda_l)]
+L\;[\theta(\lambda_j)-\theta(\lambda_k)].
\end{equation}
We already know the function $\theta(\lambda)$ increases monotonically, the sign of each term in this equation should be the same. Thus if $n_j>n_k$, then $\lambda_j>\lambda_k$, and if $n_j=n_k$, then $\lambda_j=\lambda_k$.

The above relation yields the intervals of different solutions $\lambda_j$ satisfy inequality
\begin{equation}
|\lambda_j-\lambda_k|\geq\frac{2\pi(n_j-n_k)}{{2L\over{\kappa}}(1+D)}
\geq\frac{2\pi}{{2L\over{\kappa}}(1+D)};\quad j\neq k
\end{equation}
where $D=N/L$ is the density of holomorphic QCD particles, and one also uses the inequality
\begin{equation}
0<K(\lambda,\mu)<\frac{2}{\kappa};\quad \mbox{Im}\lambda=\mbox{Im}\mu=0.
\end{equation}

Let us define a function $\lambda(x)$ ($x\in\mathbb{R}^1$), which is closely related to the solution of the Bethe equations (\ref{Log-Bethe-eq}),
\begin{equation}
2\pi Lx=\sum^N_{j=1}\theta(\lambda(x)-\lambda_j)+L\;{\theta}(\lambda(x)).
\label{Beq-coordinate}\end{equation}
According to analysis of Yang-Yang action, we know that $\lambda(x)$ does exist and is a monotonically increasing function of $x=n_j/L$.
The function value $\lambda(n_j/L)$ is just the corresponding $\lambda_j$ from the solution of (\ref{Log-Bethe-eq}),
\begin{equation}
\lambda\left({n_j\over{L}}\right)=\lambda_j,\quad \lambda_j\in\{\lambda_k\}.
\end{equation}
The values of $\lambda_j$ can be regard as the momenta of the particles, and taking the number $m\notin\{n_j\}$ ($m$ is integer for odd $N$ and half-integer for even $N$), the corresponding function value of $\lambda(x)$
\begin{equation}
\lambda_m=\lambda\left({m\over{L}}\right),
\end{equation}
 is called the momentum of the hole.
So each number $m$ (integer or half-integer), defines a vacancy. A filled vacancy is a particle, and a free vacancy is a hole.
The total number of particles and holes gives the complete number of vacancies. The density of vacancies defined as $\rho_t(\lambda)=\rho_t(\lambda(x))=dx(\lambda)/d\lambda$.
Differentiating the above equation (\ref{Beq-coordinate}) with respect to spectral parameter $\lambda$, one has
\begin{equation}
2\pi\rho_t(\lambda)={1\over{L}}\sum^{N}_{j=1}K(\lambda(x),\lambda_j)+K(\lambda(x)).
\end{equation}
Change the sum to an integral, one has the linear integral equation for $\rho_t(\lambda)$
\begin{equation}
2\pi\rho_t(\lambda)=\int^{+\infty}_{-\infty}K(\lambda,\mu)\rho_p(\mu)d\mu+K(\lambda).
\label{rho-t-int}\end{equation}
Here the number (density) of vacancies can be simply expressed as the sum of the numbers of particles and holes
\begin{equation}
\rho_t(\lambda)=\rho_p(\lambda)+\rho_h(\lambda).
\end{equation}

\section*{Appendix C: Existence of solution to Yang-Yang equation}
\setcounter{equation}{0}
\renewcommand{\theequation}{C.\arabic{equation}}

\textbf{Theorem 3.} A solution of the Yang-Yang equation does exist.
\\
Proof: To start with, we construct the following functions:
\begin{equation}
\varepsilon_0(\lambda)=\frac{-2}{\lambda^2+1}-h;
\end{equation}
\begin{equation}
\varepsilon_{m+1}(\lambda)=\frac{-2}{\lambda^2+1}-h+A_m;
\label{epsilon-m+1}\end{equation}
\begin{equation}
A_m=-\frac{T}{2\pi}\int^{+\infty}_{-\infty}K(\lambda,\mu)\ln(1+e^{-{\varepsilon_m(\mu)/T}})d\mu.
\label{Am}\end{equation}
The sequence of the series function $\varepsilon_j(\lambda)$ decreases at every point $\lambda$,
\begin{equation}
\varepsilon_0(\lambda)>\varepsilon_1(\lambda)>\cdots\varepsilon_m(\lambda)>\varepsilon_{m+1}(\lambda)>\cdots
\label{seq-decrease}\end{equation}
and we suppose it has a lower bound, which we will prove below:
\begin{equation}
\varepsilon_m(\lambda)\geq\frac{-2}{\lambda^2+1}+x_0+2,
\label{epsilon-lower-bound}\end{equation}
where $x_0$ is some constant. These properties mean that the limit
$\displaystyle\varepsilon(\lambda)=\lim_{m\rightarrow\infty}\varepsilon_m(\lambda)$
does exist and is a solution of the Yang-Yang equation.
The decrease property (\ref{seq-decrease}) can be obtained from the fact that $A_m<0$ and $\delta A_m/\delta\varepsilon_m>0$.
In order to prove that $\varepsilon_m(\lambda)$ is bounded from below, we should know that $\varepsilon_m(\lambda)$ is an even function, and
\begin{equation}
\varepsilon_m(\lambda_2)>\varepsilon_m(\lambda_1),\quad \text{if}\,\lambda_2>\lambda_1\geq 0.
\label{increase-positive-axis}\end{equation}
This property can be proved by induction in $m$.
Supposing that the above equation (\ref{increase-positive-axis}) is true for some $\varepsilon_m$, we prove that it is also true for $\varepsilon_{m+1}$. Indeed, the first order derivative of $\varepsilon_{m+1}$ has the form,
\begin{eqnarray}
\varepsilon'_{m+1}(\lambda)&=&\frac{4\lambda}{(\lambda^2+1)^2}+\frac{1}{2\pi}
\int^{+\infty}_{-\infty}
K(\lambda,\mu)\frac{\varepsilon'_{m}(\mu)}{1+e^{{\varepsilon_m(\mu)/T}}}d\mu\nonumber\\
&=&\frac{4\lambda}{(\lambda^2+1)^2}+\frac{1}{2\pi}
\int^{+\infty}_{0}
[K(\lambda,\mu)-K(\lambda,-\mu)]\frac{\varepsilon'_{m}(\mu)}{1+e^{{\varepsilon_m(\mu)/T}}}d\mu>0.
\end{eqnarray}
Since $K(\lambda,\mu)>K(\lambda,-\mu)$ at $\lambda>0$ and $\mu>0$, thus we have proved (\ref{increase-positive-axis}).

As the term $A_m$ (\ref{Am}) is also an even function, which increases monotonically on the positive semi-axis, one has
\begin{equation}
\varepsilon_m(\lambda)\geq\frac{-2}{\lambda^2+1}+2+\varepsilon_m(0).
\label{epsilon-bound}\end{equation}
Based on the above properties (\ref{epsilon-m+1}) and (\ref{epsilon-bound}), one can write the following inequality
\begin{equation}
\varepsilon_{m+1}(\lambda)\geq-2-h-\frac{T}{2\pi}\int^{+\infty}_{-\infty}
K(0,\mu)\ln(1+e^{-[\varepsilon_m(0)-\frac{2}{\mu^2+1}+2]/T})d\mu.
\label{inequal}\end{equation}
Define the function
\begin{eqnarray}
f(x)&\equiv&-2-h-\frac{T}{2\pi}\int^{+\infty}_{-\infty}
K(0,\mu)\ln(1+e^{[\frac{2}{\mu^2+1}-2-x]/T})d\mu\nonumber\\
&=&-2-h+x-\frac{T}{2\pi}\int^{+\infty}_{-\infty}
K(0,\mu)\ln(e^{x/T}+e^{[\frac{2}{\mu^2+1}-2]/T})d\mu.
\end{eqnarray}
Then one can rewrite (\ref{inequal}) as follows
\begin{equation}
\varepsilon_{m+1}(0)\geq f(\varepsilon_{m}(0)).
\label{epsilon-ineq}\end{equation}
The function $f(x)$ increases monotonically, and $f(x)<-h-2$. While the function $f(x)-x$ decreases monotonically in the real axis $[-\infty,+\infty]$.
Thus the equation
\begin{equation}
f(x_0)-x_0=0
\end{equation}
possesses a unique solution. At fixed temperature, the constant $x_0$ is uniquely defined by $h$, and vice versa,
\begin{equation}
h=-2-\frac{T}{2\pi}\int^{+\infty}_{-\infty}
K(0,\mu)\ln(e^{x_0/T}+e^{[\frac{2}{\mu^2+1}-2]/T})d\mu.
\label{h}\end{equation}
So $h$ is a monotonically decreasing function of $x_0$ in the interval $[-\infty,+\infty]$.
Now one can prove that
\begin{equation}
\varepsilon_{m}(0)\geq x_0,\quad m\geq0.
\label{epsilon-prove}\end{equation}
First one notes that
\begin{equation}
\varepsilon_{0}(0)=-2-h>f(x_0)=x_0.
\end{equation}
Then one uses induction in $m$. Supposing $\varepsilon_{m}(0)\geq x_0$, from (\ref{epsilon-ineq}) and the monotonicity of the function $f(x)$, one has
\begin{equation}
\varepsilon_{m+1}(0)\geq f(\varepsilon_{m}(0))\geq f(x_0)=x_0.
\end{equation}
Thus (\ref{epsilon-prove}) is proved. Together with (\ref{epsilon-bound}) one arrives at (\ref{epsilon-lower-bound}), which proves the existence of the solution to the Yang-Yang equation.


\begin{thebibliography}{99}

\bibitem{Lipatov94} L.N.Lipatov, {\it ``High energy asymptotics of multi-color QCD and exactly solvable lattice models''\/} JETP Lett. 59 (1994) 596-599; Pisma Zh.Eksp.Teor.Fiz. 59 (1994) 571-574.

\bibitem{Faddeev-Korchemsky1995} L. D. Faddeev, G. P. Korchemsky, {\it Phys. Lett.} {\bf B 342} (1995) 311-322.

\bibitem{ABA}
L.D. Faddeev, {\it ``Algebraic aspects of Bethe ansatz''\/},
 Stony Brook preprint, ITP-SB-94-11, Mar 1994; hep-th/9404013
 {\it ``The Bethe ansatz''\/}, Andrejewski lectures, Freie Univ.
 preprint, SFB-288-70, Jun 1993;
 {\it ``Lectures on quantum inverse scattering method''\/}
 in Nankai Lectures on Mathematical Physics,
 Integrable Systems, ed. by X.-C.Song, pp.23-70,
 Singapore: World Scientific, 1990.

\bibitem{Korepin93}
V.\,E. Korepin, N.\,M. Bogoliubov and A.\,G. Izergin, {\it Quantum Inverse Scattering Method and Correlation Functions}, Cambridge University Press, 1993.

\bibitem{Kharzeev:2017qzs}
  D.~E.~Kharzeev and E.~M.~Levin,
  Phys.\ Rev.\ D {\bf 95}, no. 11, 114008 (2017)
  doi:10.1103/PhysRevD.95.114008
  [arXiv:1702.03489 [hep-ph]].

\bibitem{Izergin-Korepin1982} A. G. Izergin and V. E. Korepin, Doklady Akademii Nauk vol 259, (l981), page 76.


\bibitem{Korepin-Izergin1982} V. E. Korepin and A.G. Izergin, {\it ``Lattice model connected with nonlinear Schrodinger equation''\/}, {\it Sov. Phys. Doklady}, 26 (1981) 653-654;
{\it ``Lattice versions of quantum field theory models in two dimensions''\/}, {\it Nucl. Phys.} {\bf B 205} (1982) 401-413.



\bibitem{Yang1968} C.~Yang and C.~Yang, {\em J.Math.Phys.} {\bf 10} (1969) 1115--1122.


\bibitem{Faddeev1980} L.D. Faddeev, {\it Soviet Sci. Reviews, Contemporary Math. Phys.},
{\bf C1} (1980) 107.

\bibitem{Tarasov-Takhtajan-Faddeev1983} V.O.Tarasov, L.A.Takhtajan and L.D.Faddeev, {\it ``Local hamiltonians for integrable quantum models on a lattice''\/}, {\it Theor. Math. Phys.}
 {\bf 57} (1983) 163-181.




\bibitem{Faddeev}  L.D. Faddeev and V.E. Korepin, Quantum Theory of Solitons,  Physics Reports vol 42 (1), pages 1-87,  1978.





\bibitem{Essler2014} Lars Bonnes, Fabian H. L. Essler, and Andreas M. Lauchli, ``Light-Cone" Dynamics After Quantum Quenches in Spin Chains, Phys. Rev. Lett. 113, 187203,(2014).


\bibitem{Bose2018} Abolfazl Bayat, Bedoor Alkurtass, Pasquale Sodano, Henrik Johannesson, and Sougato Bose, Measurement Quench in Many-Body Systems, Phys. Rev. Lett. 121, 030601,(2018).


\bibitem{Izergin-Korepin-Reshetikhin-1989} A. G. Izergin, V. E. Korepin and N. Y. Reshetikhin, Conformal dimensions in Bethe ansatz solvable models, J. Phys. A: Math. Gen. 22, 2615-2620,(1989).


\bibitem{Bogoliubov1987} N.M. Bogoliubov, A.G. Izergin, and N.Yu. Reshetikhin, Finite-size
corrections and infrared asymptotics of the correlation functions in
two dimensions, J. Phys. A20, 5361-9, (1987).

\bibitem{Korepin2004} V. E. Korepin, Universality of Entropy Scaling in One Dimensional Gapless Models, Phys. Rev. Lett. 92, 096402, (2004).





\bibitem{Gorsky02} A.~Gorsky, I.~I.~Kogan and G.~Korchemsky,
  ``High energy QCD: Stringy picture from hidden integrability,''
  JHEP {\bf 0205}, 053 (2002). arXiv:hep-th/0204183.

\end{thebibliography}
\end{document}